\documentclass{llncs}
\usepackage{makeidx}  
\usepackage{graphicx}
\usepackage{caption}
\usepackage{subfig}
\usepackage{amssymb,amsmath,amsfonts}
\usepackage{xcolor}
\usepackage[title]{appendix}

\begin{document}
%
%
%
%
%
\title{On Profiling Bots in Social Media}
%
%
\author{Richard J. Oentaryo  \and
         Arinto Murdopo \and
         Philips K. Prasetyo \and
         Ee-Peng Lim
}

\tocauthor{Richard Oentaryo, Arinto Murdopo, Philips Prasetyo, Ee-Peng Lim}
\institute{Living Analytics Research Centre, Singapore Management University\\
\email{\{roentaryo, arintom, pprasetyo, eplim\}@smu.edu.sg}}

\maketitle              

\begin{abstract}
The popularity of social media platforms such as Twitter has led to the proliferation of automated bots, creating both opportunities and challenges in information dissemination, user engagements, and quality of services. Past works on profiling bots had been focused largely on malicious bots, with the assumption that these bots should be removed. In this work, however, we find many bots that are benign, and propose a new, broader categorization of bots based on their behaviors. This includes \emph{broadcast}, \emph{consumption}, and \emph{spam} bots. To facilitate comprehensive analyses of bots and how they compare to human accounts, we develop a systematic profiling framework that includes a rich set of features and classifier bank. We conduct extensive experiments to evaluate the performances of different classifiers under varying time windows, identify the key features of bots, and infer about bots in a larger Twitter population. Our analysis encompasses more than 159K bot and human (non-bot) accounts in Twitter. The results provide interesting insights on the behavioral traits of both benign and malicious bots.

\end{abstract}

\section{Introduction}
\label{sec:intro}

In recent years, we have seen a dramatic growth of people's activities taking place in social media. Twitter, for example, has evolved from a personal microblogging site to a news and information dissemination platform. The openness of the Twitter platform, however, has made it easy for a user to set up an automated social program called \emph{bot}, to post tweets on his/her behalf.

The proliferation of bots has both good and bad consequences \cite{Chu2012,Ferrara2014}. On the one hand, bots can generate benign, informative tweets (e.g., news and blog updates), which enhance information dissemination. Bots can also be helpful for the account owners, e.g., bots that aggregate contents from various sources based on the owners' interests. On the other hand, spammers may exploit bots to attract regular accounts as their followers, enabling them to hijack search engine results or trending topics, disseminate unsolicited messages, and entice users to visit malicious sites \cite{Ghosh2012,Hu2013,Ferrara2014}. In addition to deteriorating user experience and trust, malicious bots may cause more severe impacts, e.g., creating panic during emergencies, biasing political views, or damaging corporate reputation \cite{Wang2010,Ferrara2014}.

It is thus important to characterize different types of bots and understand how they compare with human users. Recent studies have shown the importance of profiling bots in social media \cite{Wang2010,Stringhini2010,Lee2011,Chu2012,Hwang2012,Wagner2012,Ghosh2012,Hu2013,Boshmaf2013,Ferrara2014,Abokhodair2015,Subrahmanian2016}, but these works have focused mainly on malicious (e.g., spam) bots, failing to account for other types of benign bots. With the rise of new services and intelligent apps in Twitter, benign bots are increasingly becoming prominent as well. 

Comprehensive profiling of both malicious and benign bots would offer several major benefits. In information dissemination and retrieval, knowing the activity traits of both bot types and the nature of their tweet contents can improve search and recommendation services by separating tweets of bots from those of humans, returning more relevant, personalized search results, and promoting certain products/services more effectively. For social science research, a more accurate understanding of human interactions and information diffusion patterns \cite{Ferrara2014,Freitas2014} can also be obtained by filtering out activity biases generated by bots. In turn, these would benefit the overall user community as well.

\begin{figure*}[!t]
\centering
\includegraphics[width=1.0\textwidth]{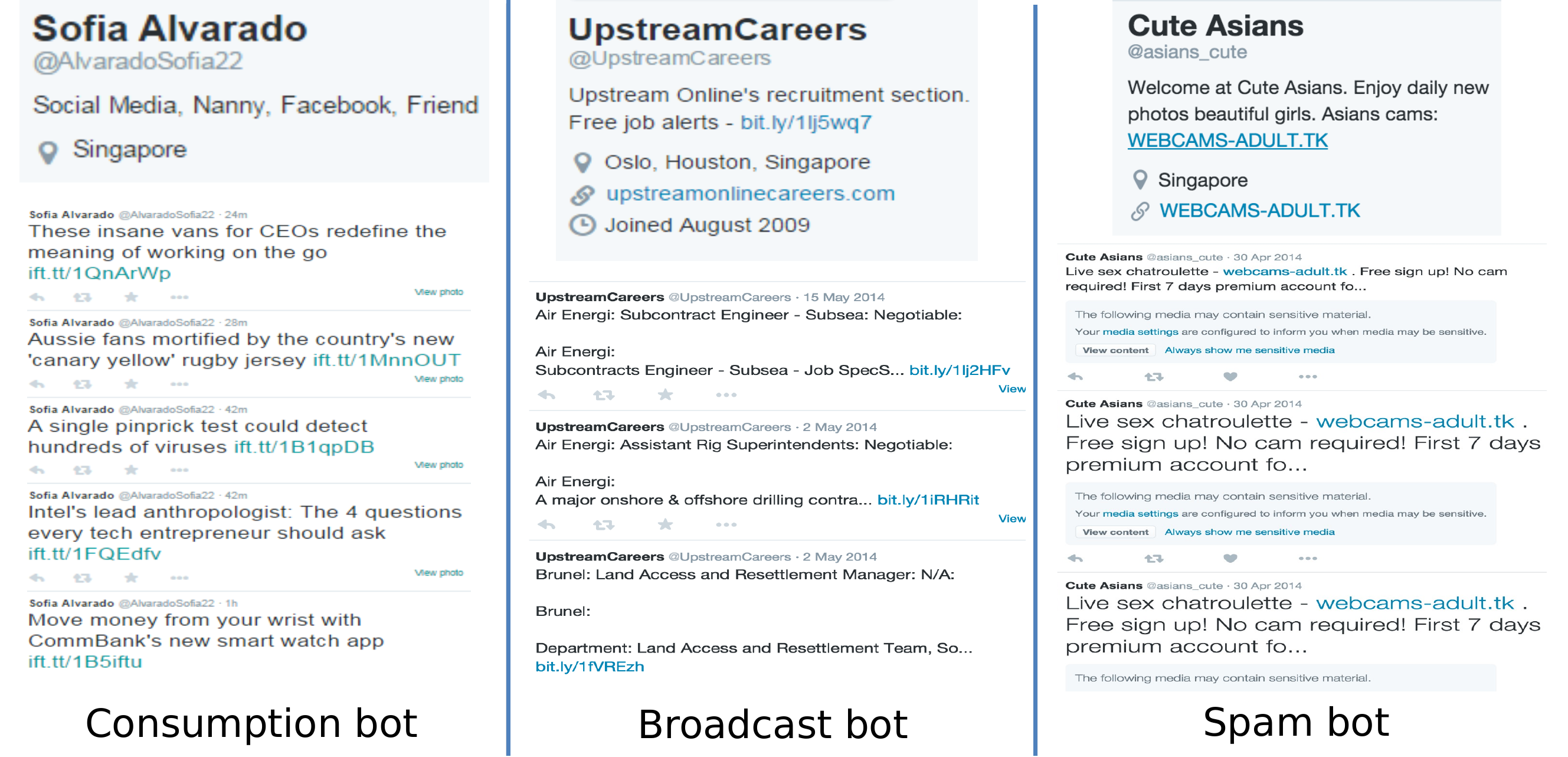}
\caption{Examples of broadcast, consumption and spam bots in Twitter}
\vspace{-4mm}
\label{fig:examples}
\end{figure*}

To illustrate the usefulness of profiling bots, consider the examples in Fig. \ref{fig:examples}, of different types of benign and malicious bots (which we further describe in Section \ref{sec:definition}). The first example is a user who utilizes the IFTTT service\footnote{\scriptsize https://ifttt.com} to gather contents from diverse sources for her own consumption. Knowing that she uses a consumption bot, Twitter can provide a new service to organize the unstructured contents, or recommend new contents that match her interest. The second example involves a broadcast bot managed by a job agency to advertise job openings. Twitter recently introduced a new feature called \emph{promoted tweets}\footnote{\scriptsize https://business.twitter.com/solutions/promoted-tweets} and, knowing it is a (benign) broadcast bot, Twitter can recommend the feature to help the agency reach a wider audience. The last example shows a malicious, spam bot that lures users to visit adult websites, posssibly containing harmful malware. For such a bot, Twitter may develop a strategy to demote---or even filter out---its posts, so that the followers do not see them on their tweet streams.

\textbf{Contributions}. In this paper, we present a new categorization of bots based on long-term observations on the behaviors of various automated accounts in Twitter. To our best knowledge, this work is the first extensive study on both \emph{benign} and \emph{malicious} Twitter bots, with detailed analyses on both their static and dynamic patterns of activity. In recent years, Twitter bots have evolved rapidly, and so our work also provides a more timely study that offers updated insights on the bot characteristics.
Our findings should also benefit social science and network mining researches. We summarize our key contributions below:

\begin{itemize}
\item We propose a new categorization of Twitter bots based on their behavioral traits. In contrast to past studies that focus largely on malicious bots, our study encompasses more detailed examinations of both malicious and benign bots, as well as how they compare to human accounts. For this, we have studied a large dataset of more than 159K Twitter accounts, out of which we have manually labeled 1.6K bot and human accounts. 

\item To facilitate comprehensive analyses on bots, we develop a systematic profiling framework that includes a rich set of numeric, categorical, and series features. This enables us to examine both the static and dynamic patterns of bots, which span various user profile, tweet, and follow network entities. Our framework also features a classifier bank that includes prominent classification algorithms, thus allowing us to comprehensively evaluate various algorithms so as to identify the best approach for bot profiling.

\item We carry out extensive empirical studies to evaluate the performance of our classifiers under different time windows and to identify the most relevant, discriminating features that characterize both benign and malicious bots. We also conduct a novel study to assess the generalization ability of our method on unseen, unlabeled Twitter accounts, based on which we infer the behavioral traits of bots in a larger Twitter population.
\end{itemize}


\section{Background and Related Work}
\label{sec:related}

A number of studies have been conducted to identify and profile bots in social media. To detect spam bots, Wang \cite{Wang2010} utilized content- and graph-based features, derived from the tweet posts and follow network connectivity respectively. Chu \emph{et al.} \cite{Chu2012} investigated whether a Twitter account is a human, bot, or cyborg. Here a bot was defined as an aggresive or spammy automated account, while cyborg refers to a bot-assisted human or human-assisted bot. Different from our work, the bots defined in \cite{Chu2012} are more of malicious nature, and the study did not provide further categorization/analysis of benign and malicious bots in Twitter.

To investigate on spam bots, Stringhini \emph{et al.} \cite{Stringhini2010} created honey profiles on Facebook, Twitter and MySpace. By analyzing the collected data, they identified anomalous accounts who contacted the honey profiles and devised features for detecting spam bots. Going further, Lee \emph{et al.} \cite{Lee2011} conducted a 7-month study on Twitter by creating 60 social honeypots that try to lure ``content polluters'' (a.k.a. spam bots). Users who follow or message two or more honeypot accounts are automatically assumed to be content polluters. There are also related works on spam bot detection based on social proximity \cite{Ghosh2012} or both social and content proximities \cite{Hu2013}. Tavares and Faisal \cite{Tavares2013} distinguished between personal, managed, and bot accounts in Twitter, according to their tweet time intervals.

Ferrara \emph{et al.} \cite{Ferrara2014} built a web application to test if a Twitter account behaves like a bot or human. They used the list of bots and human accounts identified by \cite{Lee2011}, and collected their tweets and follow network information. This study, however, covers only malicious bots. Dickerson \emph{et al.} \cite{Dickerson2014} used network, linguistic, and application-oriented features to distinguish between bots and humans in the 2014 Indian election. Abokhodair \emph{et al.} \cite{Abokhodair2015} studied on a network of bots that collectively tweet about the 2012 Syrian civil war. This study covers both malicious (e.g., phishing) and benign (e.g., testimonial) bots. In contrast to our work, however, their findings are tailored to a specific event (i.e., the civil war) and may not be applicable to other bot types in a larger Twitter population.

There are also studies aiming to quantify the susceptibility of social media users to the influence of bots \cite{Hwang2012,Wagner2012,Boshmaf2013}. By embedding their bots into the Facebook network, Boshmaf \emph{et al.} \cite{Boshmaf2013} demonstrated that users are vulnerable to phishing (e.g., exposing their phone number or address). The susceptibility of users is also evident in Twitter \cite{Hwang2012,Wagner2012}. Freitas \emph{et al.} \cite{Freitas2014} tried to reverse-engineer the infiltration strategies of malicious Twitter bots in order to understand their functioning.
Most recently, Subrahmanian \emph{et al.} \cite{Subrahmanian2016} reported the winning solutions of the DARPA Twitter Bot Detection Challenge. Again, however, all these studies deal mainly with malicious bots and ignore benign bots.

\section{New Categorization of Bots}
\label{sec:definition}

We define a bot as a Twitter account that generates contents and interacts with other users automatically---at least according to human judgment. Our definition thus includes \emph{both} benign and malicious bots. Based on long-term observations on Twitter data, we propose to categorize Twitter bots into three main types: 

\begin{figure}[!t]
\centering
\includegraphics[width=0.97\columnwidth]{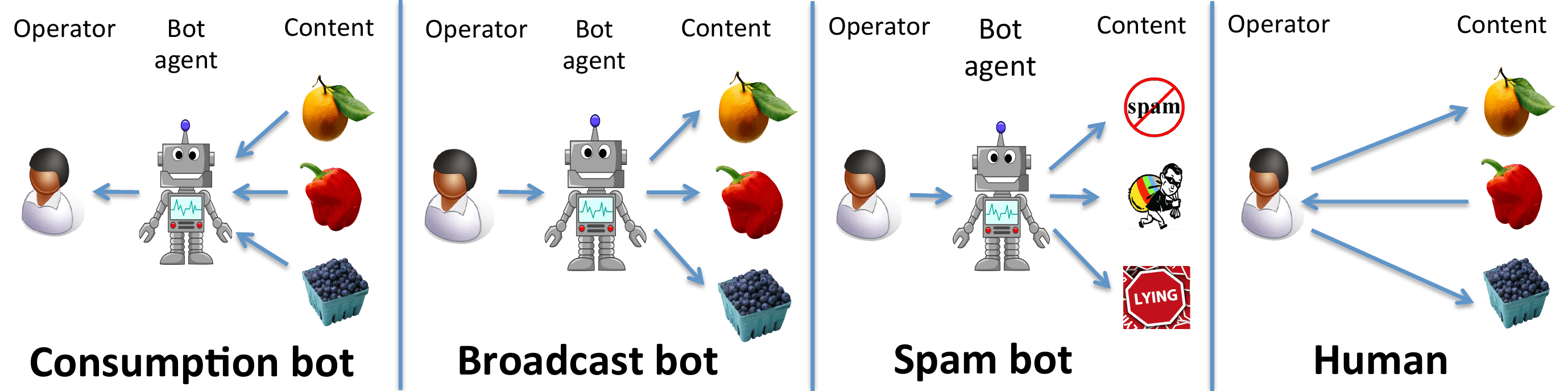}
\vspace{-2mm}
\caption{Bot and human accounts in Twitter}
\vspace{-4mm}
\label{fig:cartoon}
\end{figure}

\begin{itemize}
\item \textbf{Broadcast bot}. This bot aims at disseminating information to general audience by providing, e.g., benign links to news, blogs or sites. Such bot is often managed by an organization or a group of people (e.g., bloggers).

\item \textbf{Consumption bot}. The main purpose of this bot is to aggregate contents from various sources and/or provide update services (e.g., horoscope reading, weather update) for personal consumption or use.

\item \textbf{Spam bot}. This type of bots posts malicious contents (e.g., to trick people by hijacking certain account or redirecting them to malicious sites), or promotes harmless but invalid/irrelevant contents aggressively.
\end{itemize}

Fig. \ref{fig:cartoon} illustrates the three bot types, where the arrow direction represents the flow of information. It is worth noting that our proposed categorization is more general than the taxonomy put forward in \cite{Mitter2013}, which covers mainly malicious bots. Our categorization is also general enough to cater for new, emerging types of bot (e.g., chatbots can be viewed as a special type of broadcast bots).

\section{Dataset}
\label{sec:data}

\textbf{Data collection}.
Our study involves a Twitter dataset generated by users in Singapore and collected from 1 January to 30 April 2014 via the Twitter REST and streaming APIs\footnote{\scriptsize https://dev.twitter.com/overview/}. Starting from popular seed users (i.e., users having many followers), we crawled their follow, retweet, and user mention links. We then added those followers/followees, retweet sources, and mentioned users who state Singapore in their profile location. With this, we have a total of 159,724 accounts.

\begin{table}[!t]
\scriptsize
\centering
\caption{Distribution of our Twitter dataset}
\begin{tabular}{|c|c|c|c|c|}
\hline
\multicolumn{4}{|c|}{\textbf{Labeled data}} & \textbf{Unlabeled data} \\
\cline{1-4}
\emph{Consumption bot} & \emph{Broadcast bot} & \emph{Spam bot} & \emph{Human account} & \\
\hline
313 & 171 & 105 & 1,024 & 158,111 \\ 
\hline
\multicolumn{5}{l}{Total no. of labeled data = 1,613; Total no. of data = 159,724}
\end{tabular}
\label{tab:label_dist}
\vspace{-2mm}
\end{table}

\begin{figure*}[!t]
\scriptsize
\centering
\begin{tabular}{@{}c@{}}
\includegraphics[width=0.45\columnwidth]{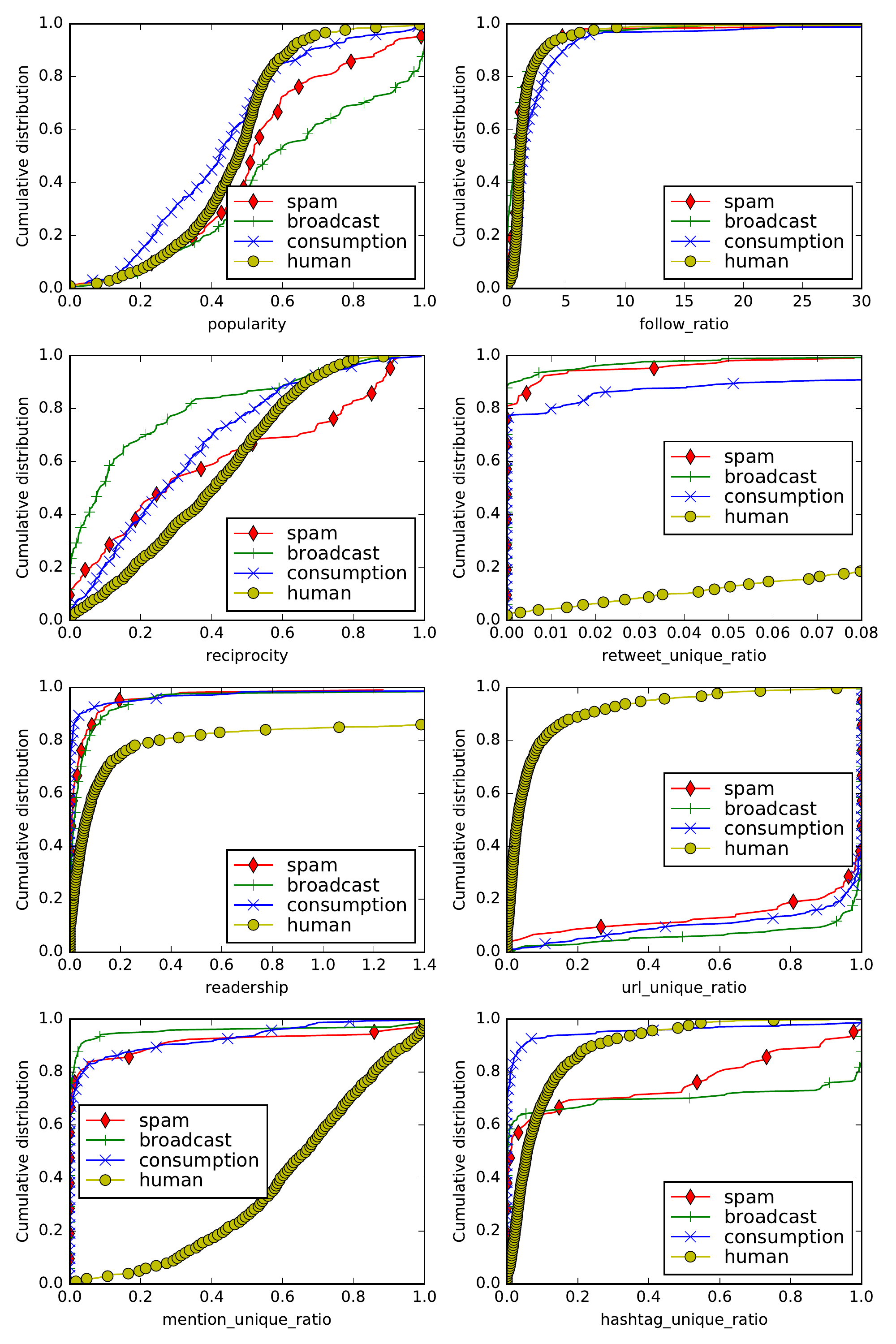}\\
(a) Cumulative distribution functions
\end{tabular}
\begin{tabular}{@{}c@{}}
\includegraphics[width=0.54\columnwidth]{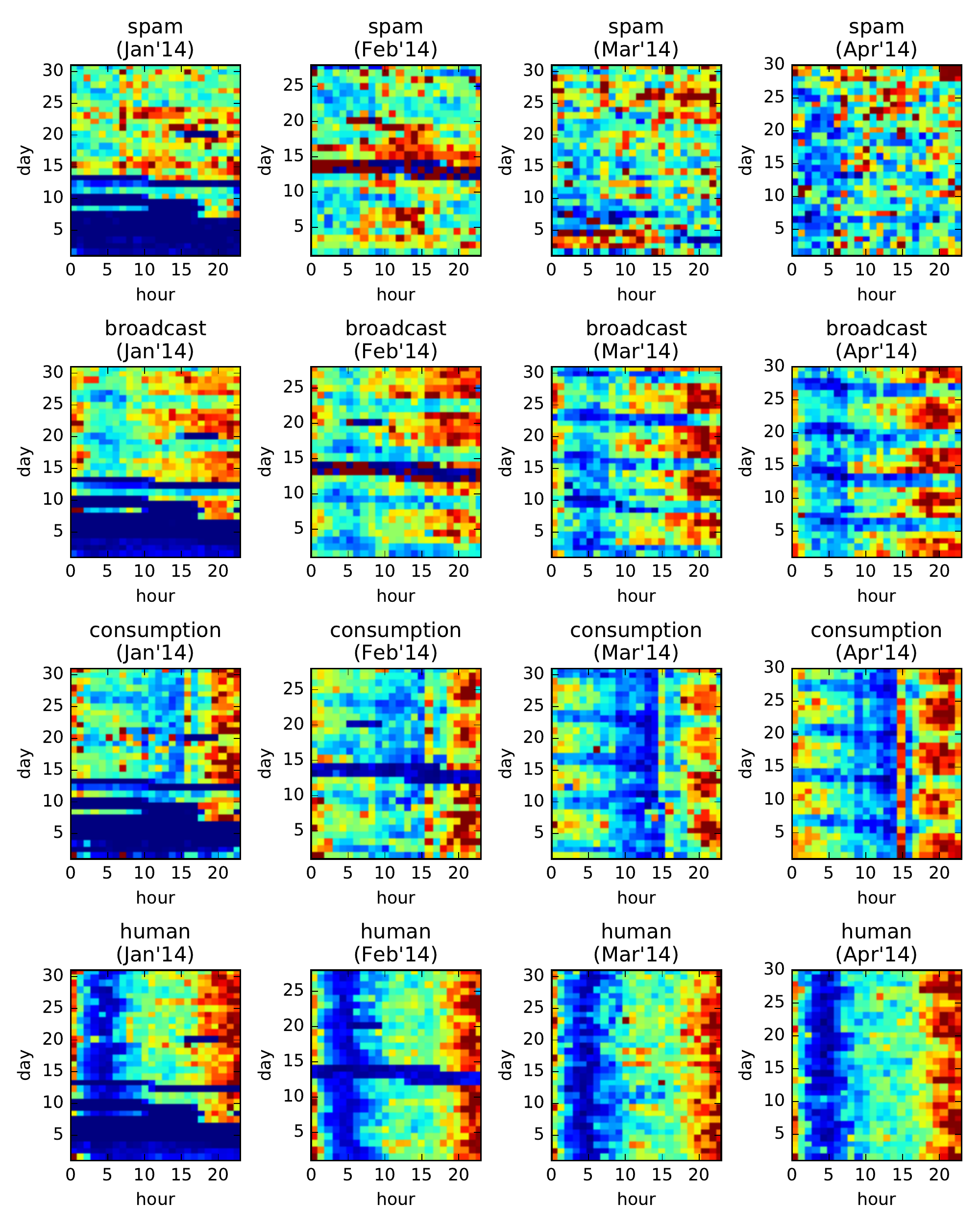}\\
(b) Temporal dynamics
\end{tabular}
\caption{Statistics of humans and bots in our labeled Twitter data}
\vspace{-2mm}
\label{fig:results}
\end{figure*}

To identify bots, we first checked active accounts who tweeted at least 15 times within the month of April 2014. 
We then manually labeled these accounts and found 589 bots. As many more human users are expected in the Twitter population, we randomly sampled the remaining accounts, manually checked them, and identified 1,024 human accounts. In total, we have 1,613 labeled accounts, as summarized in Table \ref{tab:label_dist}. The labeling was done by four volunteers, who were carefully instructed on the definitions in Section \ref{sec:definition}. The volunteers agree on more than $90\%$ of the labels, and any labeling differences in the remaining accounts are resolved by consensus. Also, if an account exhibits both human and bot characteristics, we determine the label based on the majority posting patterns.

\textbf{Exploratory analysis}. 
We conducted a preliminary study on our 1,613 labeled data to get a glimpse of the activity patterns of bots as well as human accounts. Fig. \ref{fig:results}(a) shows the cumulative distribution functions (CDF) of several key attributes. An early increase in CDF value means a more skewed distribution. We focus on key attributes that reflect a user's social and posting patterns: $popularity = \frac{|F|}{|E| + |F|}$, $follow\_ratio = \frac{|E|}{|F|}$, $reciprocity = \frac{|E \cap F|}{|E \cup F|}$, $retweet\_unique\_ratio = \frac{|R|}{|T|}$, $url\_unique\_ratio = \frac{|U|}{|T|}$, $mention\_unique\_ratio = \frac{|M|}{|T|}$, $hashtag\_unique\_ratio = \frac{|H|}{|T|}$, where $E$, $F$, $R$, $T$, $U$, $M$, $H$ are the set of followees, followers, retweets, tweets, URLs, user mentions, and hashtags for a given account, respectively. We also define $readership = \frac{retweeted}{|T|}$, where $retweeted$ is the number of times a user's tweets get retweeted (by others). Fig. \ref{fig:results}(b) shows heatmaps of tweet counts $|T|$ for different days and hours over 4 months.

\emph{How do humans compare with bots and how do bots differ from one another}? 
The $popularity$, $follow\_ratio$, and $reciprocity$ results in Fig. \ref{fig:results}(a) suggest that bots (except for consumption bots) generally have more followers than followees, but are less reciprocal (i.e., follow each other) than humans. Based on the $retweet\_unique\_ratio$ and $readership$ results, humans are more likely to reshare contents from others and have their contents reshared than bots, respectively. Similarly, the $mention\_unique\_ratio$ result suggests that humans are more likely to mention (i.e., talk to) others than bots. Meanwhile, the $url\_unique\_ratio$ and $hashtag\_unique\_ratio$ results show the bots tend to include more diverse web links and topics than humans, respectively. Finally, comparisons among the three bot types show that broadcast bots are the most popular and post the most diverse URLs and hashtags, but they are the least reciprocal and rarely mention others. A plausible reason is that broadcast bots are typically used by organizations solely for information dissemination, and not for interaction with others.

\emph{How do activities of humans and bots change over time?} 
Fig. \ref{fig:results}(b) shows that seasonality exists in the tweet activities of human and bot accounts\footnote{\scriptsize The exceptionally low tweet frequencies in the first week of January and 12-14 February are due to major downtime of our servers.}. That is, humans seldom tweet in early morning (from 2am to 7am) and post moderately from 7am to 8pm. Afterwards, their tweet traffic increases significantly between 8pm and midnight, suggesting that Singapore users are more active after dinner time and before they sleep. Meanwhile, consumption bots tweet more actively than humans from 3am to 7am (i.e., sleep hours), but are less active from 9am to 3pm (i.e., busy working/school hours). Also, consumption bots are less active in the weekends than in the weekdays. While broadcast bots have generally similar patterns to consumption bots, the former is less active during sleep hours (3am--7am) whereas the latter during busy hours (9am--3pm). We can attribute this to the intuition that broadcast bots aim to reach a wider audience during their non-sleep hours. Lastly, unlike broadcast and consumption bots, spam bots are active all days/hours, and they exhibit very random timings. In summary, different bots serve different purposes and their temporal signatures reflect these. 

\section{Profiling Framework}
\label{sec:method}

We develop a systematic profiling framework to facilitate comprehensive analyses of bots. Below we describe each component of the framework in turn.

\textbf{Database}. 
Our framework takes as input three types of database: \emph{profile}, \emph{tweet}, and \emph{follow} databases. The profile database contains user information such as the Twitter user id, screenname, location, and profile description. The tweet database contains all the tweets posted by different users, which may include various entities such as hashtags, URLs, user mentions, videos/images, retweet information, and tweet sources/devices. We collectively refer to these as \emph{tweet entities}. Finally, the follow database contains the snapshots of users' relationship network over time, which include both followers and followees of the users at different time periods. We collectively call these \emph{follow entities}.

\textbf{Feature extraction}.
This component serves to construct a \emph{feature vector} that represents a Twitter account. It takes three types of feature: \emph{numeric}, \emph{categorical}, and \emph{series}. We describe the extraction steps for each type below:

\begin{itemize}
\item For \textbf{numeric features}, we perform \emph{standarization} by scaling each feature to a unit range $[0, 1]$. This would allow us to mitigate feature scaling issues, particularly for classification methods that rely on some distance metric. Examples of numeric features are count and ratio attributes (see Table \ref{tab:features}).

\item For \textbf{categorical features}, we first select the top $K$ categories based on their frequencies in each data point, and then filter out the remaining categories. Next, we perform \emph{one-hot encoding} by transforming the top $K$ categories into a binary vector with $K$ elements. For example, a categorical attribute with four possible values: ``A'', ``B'', ``C'', and ``D'' is encoded as $[1, 0, 0, 0]$, $[0, 1, 0, 0]$, $[0, 0, 1, 0]$, and $[0, 0, 0, 1]$, respectively.

\item For \textbf{series features}, we first count the frequency of every (discrete) number in the series. For instance, given a series $[a, a, b, a, c, b, c, a, b]$, we can compute the histogram bins: $(a,4), (b,3), (c,2)$. To ensure a moderate feature size, we keep only top $100$ bins with the highest count frequencies. Subsequently, we normalize the frequencies such that they sum to 1, thus forming a probability distribution. For the previous histogram bins $(a,4)$, $(b,3)$, $(c,2)$, the normalization will result in $(a,\frac{4}{9})$, $(b,\frac{3}{9})$, $(c,\frac{2}{9})$.
\end{itemize}

\textbf{Classifier bank}. Finaly, to learn the association between the extracted features and different bot types (or human), our framework includes a classifier bank that comprises a rich collection of classification algorithms.
In our study, we employ four prominent classifiers: \emph{na\"{i}ve Bayes} (NB) \cite{Domingos1997}, \emph{random forest} (RF) \cite{Breiman2001}, and two instances of generalized linear model, i.e., \emph{support vector machine} (SVM) and \emph{logistic regression} (LR) \cite{Fan2008}.
These algorithms represent the state-of-the-art methods previously used for (malicious) bot classification. For instance, RF was utilized in \cite{Chu2012,Lee2011,Ferrara2014,Dickerson2014}, while SVM and NB were used in \cite{Wang2010,Dickerson2014}.

\section{Feature Engineering}
\label{sec:features_extracted}

We have crafted a rich set of features based on the feature extraction component in our bot profiling framework. Our feature set consists of three groups: \emph{tweet}, \emph{follow} and \emph{profile} features. For tweet features, we also distinguish between \emph{static} (i.e., time-independent) and \emph{dynamic} (i.e., time-dependent) tweet features. Table \ref{tab:features} provides a listing of all the features used in our empirical study.

\begin{table*}[!t]
\scriptsize
\centering
\caption{List of features used in our bot classification task}
\begin{tabular}{|l|l|l|}
\hline
\textbf{Group} & \textbf{Entity} & \textbf{Features}\\
\hline
Static 	 	& tweet\_word       & count (N), unique\_count (N), unique\_ratio (N), basic\_stats (N) 							\\ \cline{2-3}
tweet    	& retweet           & retweeted (N), readership (N), count (N), unique\_count (N), ratio (N),\\
features    &                   & unique\_ratio (N), basic\_stats (N)  \\ \cline{2-3}
 			& hashtag           & count (N), unique\_count (N), ratio (N), unique\_ratio (N), basic\_stats (N) 					\\ \cline{2-3}
		 	& mention           & count (N), unique\_count (N), ratio (N), unique\_ratio (N), basic\_stats (N) 					\\ \cline{2-3}
		 	& url               & count (N), unique\_count (N), ratio (N), unique\_ratio (N), basic\_stats (N) 					\\ \cline{2-3}
		 	& media             & count (N), unique\_count (N), ratio (N), unique\_ratio (N), basic\_stats (N) 					\\ \cline{2-3}
			& source            & sources (S) 																					\\
\hline
Dynamic 	& tweet             & hours (S), days (S), weekdays (S), timeofdays (S), extended\_stats (N)\\ \cline{2-3}
tweet    	& retweet           & hours (S), days (S), weekdays (S), timeofdays (S), extended\_stats (N)\\ \cline{2-3}
features  	& hashtag           & hours (S), days (S), weekdays (S), timeofdays (S), extended\_stats (N)\\ \cline{2-3}
		 	& mention           & hours (S), days (S), weekdays (S), timeofdays (S), extended\_stats (N)\\ \cline{2-3}
		 	& url               & hours (S), days (S), weekdays (S), timeofdays (S), extended\_stats (N)\\ \cline{2-3}
		 	& media             & hours (S), days (S), weekdays (S), timeofdays (S), extended\_stats (N)\\
\hline
Follow      & followees\_count  & basic\_stats (N) \\ \cline{2-3}
features    & followers\_count  & basic\_stats (N) \\ \cline{2-3}
            & mutual\_count     & basic\_stats (N) \\ \cline{2-3}
            & reciprocity       & basic\_stats (N) \\ \cline{2-3}
            & in\_reciprocity   & basic\_stats (N) \\ \cline{2-3}
            & out\_reciprocity  & basic\_stats (N) \\ \cline{2-3}
            & popularity        & basic\_stats (N) \\ \cline{2-3}
            & follow\_ratio     & basic\_stats (N) \\
\hline
Profile     & profile           & is\_geo\_enabled (C), lang (C), time\_zone (C), account\_age (N), \\
features    &                   & favourites\_count (N), listed\_count (N), statuses\_count (N), utc\_offset (N) \\
\hline
\multicolumn{3}{l}{basic\_stats: set of statistical metrics \{mean, median, min, max, std, entropy\}}\\
\multicolumn{3}{l}{extended\_stats: Cartesian product of \{timegap, hour, day, weekday, timeofday\} and basic\_stats}\\
\multicolumn{3}{l}{N: numeric feature, C: categorical feature, S: series feature}\\
\end{tabular}
\label{tab:features}
\end{table*}

\textbf{Static tweet features}. 
We generate static tweet features based on the combination of entities and statistical metrics, as shown in Table \ref{tab:features}. For instance, to generate the hashtag features of a user, we treat each hashtag as a ``bag'' and count how many times the word occurs in all of $x$'s tweets. This yields a bag-of-hashtag vector, from which we can compute first-order statistics (i.e., $count$, $unique\_count$, $mean$, $median$, $min$, and $max$) as well as second-order metrics (i.e., standard deviation ($std$) and Shannon entropy \cite{Shannon1963} ($entropy$)). We note that the second-order metrics serve to quantify the \emph{diversity} of the entities. We also compute the $ratio = \frac{count}{|T|}$ and $unique\_ratio = \frac{unique\_count}{|T|}$, where $|T|$ is the total number of tweets posted by a user. 
For the retweet entity, we additionally consider $retweeted$ and $readership$ features, as described in Section \ref{sec:data}. Finally, we consider a series feature to represent the source entity, whereby each source maps to a histogram bin containing the normalized frequency of the source.

\textbf{Dynamic tweet features}. 
For these features (cf. Table \ref{tab:features}), we introduce additional time dimensions that capture the dynamics of tweet activities, namely: \emph{hours} $\in \{0,\ldots,23\}$, \emph{days} $\in \{1,\ldots,31\}$, \emph{weekdays} $\in \{Monday,\ldots,Sunday\}$, \emph{timeofdays} $\in \{morning$ (4am--12pm), $afternoon$ (12pm--5pm), $evening$ (5pm--8pm), $night$ (8pm--4am)$\}$, and \emph{timegaps}. The timegap dimension refers to the gap (in milliseconds) between two \emph{consecutive} entity timestamps, e.g., for $N$ tweets posted by a user $x$, we can compute a timegap vector with length $(N-1)$. For each time dimension, we can then generate the series features based on the histogram binning described in Section \ref{sec:method}, as well as compute the statistical metrics such as $mean$, $median$, $min$, $max$, $std$ and $entropy$.

\textbf{Follow features}.
These features are derived by computing metrics that summarize snapshots of the follow network at different time points (cf. Table \ref{tab:features}). Let $E$ and $F$ be the set of followees and followers of a given user. In turn, we compute the $followees\_count = |E|$, $followers\_count = |F|$, $mutual\_count = |E \cap F|$. as well as ratio metrics such as $reciprocity = \frac{|E \cap F|}{|E \cup F|}$, $in\_reciprocity = \frac{|E \cap F|}{|F|}$, $out\_reciprocity = \frac{|E \cap F|}{|E|}$, $popularity = \frac{|F|}{|E| + |F|}$, and $follow\_ratio = \frac{|E|}{|F|}$. We calculate these metrics for every snapshot of the follow network at a given time point, and then compute the statistics $mean$, $median$, $min$, $max$, $std$ and $entropy$ to summarize the metrics over all time points.

\textbf{Profile features}. Finally, we also consider several basic user profile features, as per Table \ref{tab:features}.
Here, $account\_age$ refers to the lapse between the time a user first joined Twitter and the current reference time. Further details on the definitions of the other profile features can be found in \texttt{\url{https://dev.twitter.com/}}.

\vspace{-1mm}
\section{Results and Findings}
\label{sec:experiment}

This section elaborates our empirical study on bots. 
We first describe our experiment setup, and then address several research questions in Sections \ref{sec:RQ1}--\ref{sec:RQ4}.


\textbf{Evaluation metrics}. To evaluate our classifiers, we utilize three metrics popularly used in information retrieval \cite{Manning2008}: $Precision$, $Recall$ and $F1$. We report, for each class $c \in \{broadcast, consumption, spam, human\}$, the $Precision(c) = \frac{TP(c)}{TP(c) + FP(c)}$, $Recall = \frac{TP(c)}{TP(c) + FN(c)}$, and $\text{\emph{F1}(c)} = \frac{2 Precision(c) Recall(c)}{Precision(c) + Recall(c)}$, where $TP(c)$, $FP(c)$ and $FN(c)$ are the true positives, false positives, and false negatives respectively. Based on these, we also report the macro-averaged $Precision = \frac{1}{4} \sum_{c=1}^4 Precision(c)$, $Recall = \frac{1}{4} \sum_{c=1}^4 Recall(c)$, and $\text{\emph{F1}} = \frac{1}{4} \sum_{c=1}^4 \text{\emph{F1}(c)}$.

\textbf{Experiment protocols}. In this work, we consider two sets of experiment: 

\begin{itemize}
\item \textbf{Experiment} $E_1$: This set of experiment involves evaluation on our \textbf{1,613} \emph{labeled data} (see Table \ref{tab:label_dist}). For this evaluation, we use a \emph{stratified} 10-fold cross-validation (CV), whereby we split the labeled data into 10 mutually exclusive groups, each retaining the class proportion as per the original data. This stratification serves to ensure that each fold is a good representative of the whole, i.e., it retains the (unbalanced) class distribution as in the original data. For each CV iteration $f$, we then use group $f$ ($10\%$) for testing and the remaining groups $f' \neq f$ ($90\%$) for training. We report the results averaged over 10 iterations, which include $Precision(c)$, $Recall(c)$ and $F1(c)$ for each class $c$, as well as the macro-averaged $Precision$, $Recall$ and $F1$. 

\item \textbf{Experiment} $E_2$: This set of experiment serves to evaluate predictions on the remaining \textbf{158,111} \emph{unlabeled data} (see again Table \ref{tab:label_dist}). Based on this, we can infer the behavioral traits of bots in a larger Twitter population. For this experiment, we are unable to compute $Recall$, as we would have to manually verify one by one a large number of unlabeled data. Instead, we evaluate based on $Precision$ at top $K$ for each class ($K \ll$ 158,111). 
\end{itemize}

\textbf{Model parameters}. We configured our classifier bank as follows: For the NB classifier, we use the smoothing parameter $\alpha=1$. For RF, we use $N=100$ decision trees. Finally, for SVM and LR, we set the cost parameter $C=1$ and $\texttt{class\_weight}=$``balanced''; the latter is for automatically handling the imbalanced class distribution. We performed grid search to determine all these parameters, which give the optimal performances for each classifier. In particular, we varied the NB parameter from the range $\alpha \in \{0.1, 1, 10\}$. For RF, we tried $N \in \{10,20,\ldots,100\}$, and for SVM and LR, we tried $C \in \{0.01, 0.1, 1, 10, 100\}$.

\textbf{Significance test}. Finally, we use \emph{Wilcoxon signed-rank test} \cite{Wilcoxon1945} to test for the statistical significance of our results. When comparing between two performance vectors, we look at the $p$-value at a significance level of $0.01$. If the $p$-value is less than $0.01$, we say that the performance difference is indeed significant.

\subsection{How Well Can the Classifiers Predict for Bots?}
\label{sec:RQ1}

\begin{figure}[!t]
\centering
\includegraphics[width=0.6\columnwidth]{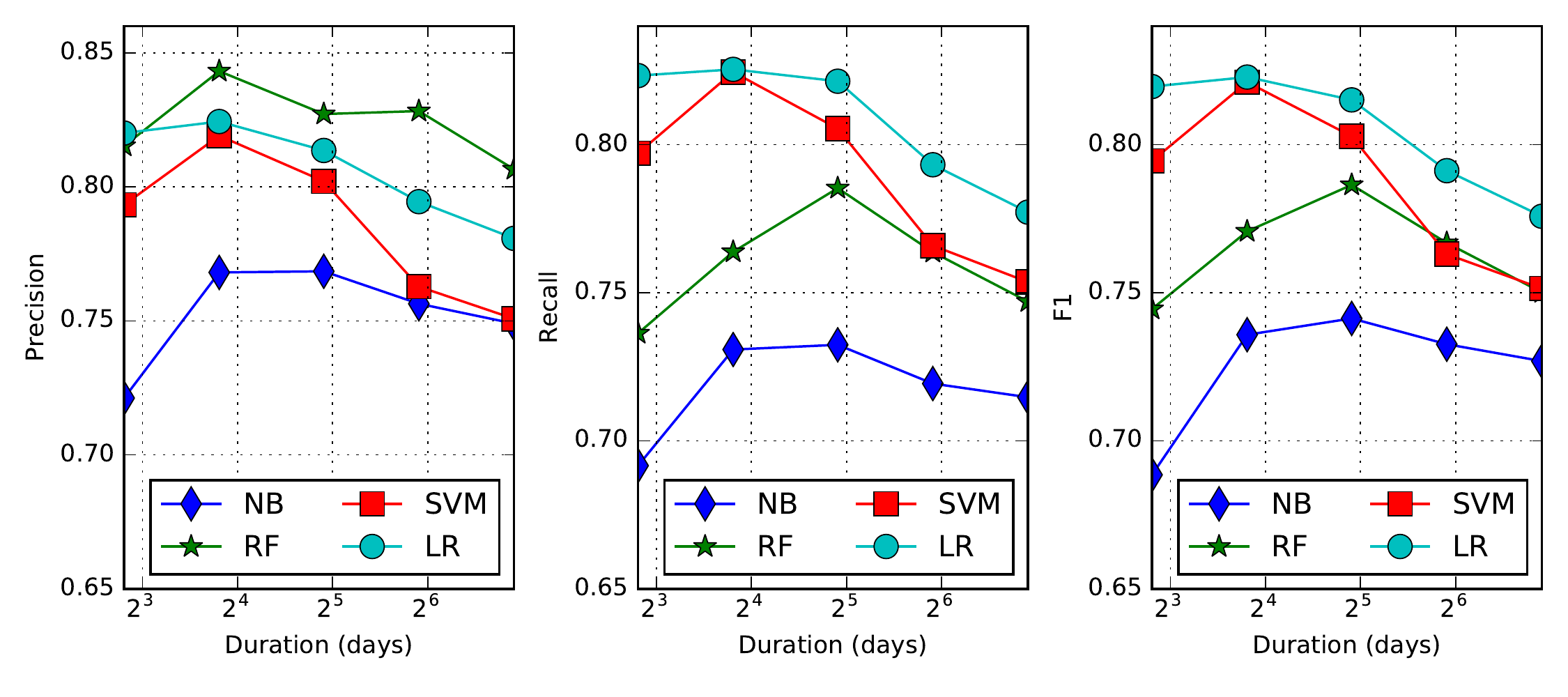}
\caption{Classification results for varying durations}
\vspace{-5mm}
\label{fig:time_window}
\end{figure}

To answer this research question, we first conduct a sensitivity study by varying the time duration for which features (cf. Table \ref{tab:features}) are generated. For this study, we use the CV procedure on our labeled data (i.e., Experiment $E_1$), whereby the classifiers were trained using all features listed in Table \ref{tab:features}. Fig. \ref{fig:time_window} shows the macro-averaged $Precision$, $Recall$, and $F1$ over 10 CV folds, with the duration varied from 1 week, 2 weeks and 1 month to 2 months and 4 months (up to 30 April 2014). Based on the $F1$ results, we can conclude that 2 weeks is the best duration and that LR outperforms the other classifiers. In this case, RF gives higher $Precision$ than LR, but its $Recall$ is much lower, and so is its $F1$. It is also shown that a tradeoff exists in choosing the duration; an overly short duration degrades the performance, which can be attributed to data scarcity. The same goes for an overly long duration, due to inclusion of outdated data.

Table \ref{tab:benchmark} shows further breakdown of the CV results for the best time duration (i.e., 2 weeks). Overall, LR and SVM give the best results, and outperform the more complex RF and simpler NB methods (except for $Precision$ of the ``spam'' class). For spam bots, RF yields higher $Precision$, but much lower $Recall$ and $F1$ than LR and SVM. While SVM and LR perform very similarly, we decided to use LR as our main classifier for two reasons: (i) LR outputs more meaningful probabilitic scores than the unbounded decision scores in SVM; and (ii) LR is more robust than SVM against variation in time duration, as we saw in Fig. \ref{fig:time_window}.

\begin{table*}[!t]
\scriptsize
\centering
\caption{Breakdown of 10-fold cross-validation results using 2-week training data}
\begin{tabular}{|l|l|rl|rl|rl|rl|rl|}
\hline
& & \multicolumn{8}{|c|}{\textbf{Class label}} & \multicolumn{2}{|c|}{\textbf{Macro}}\\
\cline{3-10}
\textbf{Metric}      & \textbf{Method} &  \multicolumn{2}{|c|}{\textbf{Broadcast}} & \multicolumn{2}{|c|}{\textbf{Consumption}} & \multicolumn{2}{|c|}{\textbf{Spam}} & \multicolumn{2}{|c|}{\textbf{Human}} & \multicolumn{2}{|c|}{\textbf{average}}\\
\hline
Precision 	& NB &          0.6519 & $(-)$ &          0.7206 & $(-)$ &          0.7069 & $(+)$ &           0.9929 &     &          0.7681 & $(-)$ \\ \cline{2-12}
			& RF &          0.5880 & $(-)$ & \textbf{0.9462} &     & \textbf{0.8636} & $(+)$ &           0.9750 & $(-)$ & \textbf{0.8432} & $(+)$ \\ \cline{2-12}
 			& SVM &\textbf{0.6952} &     &          0.9278 &     &          0.6574 & $(-)$ &  \textbf{0.9961} &     &          0.8191 & \\ \cline{2-12}
 			& LR &          0.6798 &     &          0.9366 &     &          0.6869 &     &           0.9942 &     &          0.8244 & \\
\hline
Recall 		& NB &          0.6901 & $(-)$ & \textbf{0.8818} & $(+)$ &          0.3905 & $(-)$ &           0.9609 & $(-)$ &          0.7308 & $(-)$ \\ \cline{2-12}
    		& RF & \textbf{0.8596} & $(+)$ &          0.8435 &     &          0.3619 & $(-)$ &           0.9902 &     &          0.7638 & $(-)$ \\ \cline{2-12}
    		& SVM &         0.7602 & $(-)$ &         0.8626  &     & \textbf{0.6762} & $(+)$ &  \textbf{0.9990} &     &          0.8245 & \\ \cline{2-12}
    		& LR &          0.8070 &     &          0.8498 &     &          0.6476 &     &           0.9971 &     & \textbf{0.8254} & \\
\hline
F1-score    & NB &          0.6705 & $(-)$ &          0.7931 & $(-)$ &          0.5031 & $(-)$ &           0.9767 & $(-)$ &          0.7358 & $(-)$ \\ \cline{2-12}
   			& RF &          0.6983 & $(-)$ &          0.8919 &     &          0.5101 & $(-)$ &           0.9826 & $(-)$    &          0.7707 & $(-)$ \\ \cline{2-12}
   			& SVM &         0.7263 &     & \textbf{0.8940} &     & \textbf{0.6667} &     &  \textbf{0.9976} &     &          0.8211 & \\ \cline{2-12}
		    & LR & \textbf{0.7380} &     &          0.8911 &     & \textbf{0.6667} &     &           0.9956 &     & \textbf{0.8228} & \\
\hline
\multicolumn{12}{l}{NB: n{\"a}ive Bayes, SVM: support vector machine, LR: logistic regression, RF: random forest}\\
\multicolumn{12}{l}{$(-)$: significantly worse than LR at $0.01$, $(+)$: significantly better than LR at $0.01$}
\end{tabular}
\vspace{-4mm}
\label{tab:benchmark}
\end{table*}

Based on the individual $Precision(c)$, $Recall(c)$ and $F1(c)$ of each class $c$, we can conclude that, among the bots, consumption bots are the easiest to detect, followed by broadcast and spam bots. This is expected, owing to the imbalanced class distribution as per Table \ref{tab:label_dist}. We can also compare the results of our classifiers with that of a random guess\footnote{\scriptsize Random guess w.r.t. a class $c$ refers to a classifier that assigns a proportion $p_c\%$ of the instances to class $c$, and $(1-p_c)\%$ to classes other than $c$. In this case, $Precision(c) = Recall(c) = F1(c) = p_c$, where $p_c = \frac{P(c)}{P(c)+N(c)} = \frac{TP(c) + FN(c)}{TP(c) + FN(c) + TN(c) + FP(c)}$.}. Based on the statistics in Table \ref{tab:label_dist}, the expected $F1$ scores of a random guess for broadcast bot, consumption bot, spam bot, and human classes are $10.6\%$, $19.40\%$, $6.51\%$ and $63.49\%$, respectively. Our four classifiers thus outperform the random guess baseline by a large margin.

For spam bots, several studies \cite{Lee2011,Chu2012,Ferrara2014} have reported high classification accuracies, while our results are modest by comparison, largely due to the lack of spam bot accounts in our data. However, it must be noted that these works focused largely on distinguishing between (malicious) bots vs. other accounts, whereas our study deals with a much more challenging and fine-grained categorization of broadcast, consumption and spam bots. Also, the lack of spam bots in our data can be attributed to several factors, such as our relatively strict definition of spam bot (whereby the majority of its postings need to have malicious or irrelevant contents), or our data collection process that begins with popular seed users and their connections (thus possibly missing unpopular spam bots). Nevertheless, our main focus is to analyze benign bots, which has been largely ignored in the past studies. Further studies on less prominent spam bots that post malicious contents at a sparse rate is beyond the scope of our current study.

\vspace{-2mm}
\subsection{Which Features are the Most Indicative of Each Bot Type?}
\label{sec:RQ2}

\begin{figure*}[!t]
\centering
\includegraphics[width=1.0\columnwidth]{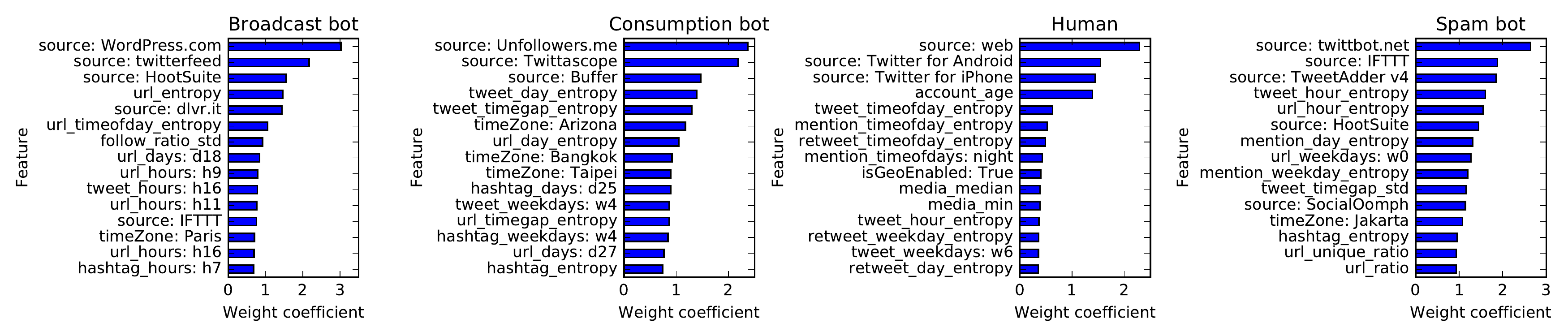}
\caption{Top discriminative features for each label in bot classification task}
\vspace{-6mm}
\label{fig:feature_importance}
\end{figure*}

In light of this research question, we trained our best classifier (i.e., LR) using all 1,613 labeled data, and look at the weight coefficients $w_{i,c}$ of each class in the trained LR. Here we use the raw weights $w_{i,c}$ instead of the absolute values $|w_{i,c}|$ or squared values $w_{i,c}^2$, as the raw weights allow us to distingush between features that correlate positively with a class label (which are our main interest) and those that correlate negatively. Fig. \ref{fig:feature_importance} shows the top 15 positively-correlated features for each class. In general, we find that the top features are dominated by the \emph{source} (i.e., where the tweets come from) and \emph{entropy-based dynamic tweet} features. Below we elaborate our feature analysis for each class further.

\textbf{Broadcast bots}. Among the top features for broadcast bots, certain sources that are popularly used for blogging (such as WordPress and Twitterfeed) or brand management (such as HootSuite) are found to be highly indicative. It is also shown that the entropy-based features for the url entity correlate strongly with broadcast bots. Recall from Section \ref{sec:features_extracted} that entropy is a second-order metric that quantifies how diverse a distribution is. Accordingly, as broadcast bots generally aim to disseminate information about certain sites/brands, we can expect that they would have more concentrated url distribution (i.e., low entropy). We will further verify this in Section \ref{sec:RQ4}. Fig. \ref{fig:feature_importance} also suggests that certain critical timings of the url postings are highly indicative of broadcast bots.

\textbf{Consumption bots}. From Fig.~\ref{fig:feature_importance}, we firstly find that the top three sources for consumption bots (i.e., Unfollowers, Twittascope, and Buffer) are service apps that allow users to track their followers/followees status, horoscope readings, and scheduled postings, respectively. Secondly, we discover that the diversity (entropy) of tweet postings is a strong indicator for consumption bots. Lastly, Fig.~\ref{fig:feature_importance} shows that certain timezones and timings (weekday and day) of the hashtag and url activities constitute yet another important set of indicators. All these led us to conclude that consumption bots post tweets in a way that follows certain timings/schedules. We will further analyze this in Section \ref{sec:RQ4}.

\textbf{Spam bots}. The result in Fig.~\ref{fig:feature_importance} suggests that there are certain sources that can be exploited by spammers to post irrelevant or unsolicited tweets. For example, TwittBot is an application that allows multiple users (and thus spammers) to post to a single Twitter account. In addition, the timing diversities of the url, mention, tweet and hashtag activities are found to be the key signatures of spam bots. As also shown in Fig. \ref{fig:results}(b) (of Section \ref{sec:data}), the temporal patterns of spam bots are highly irregular. Altogether, these suggest that spam bots have highly diverse timings (i.e., high entropy), which we will again verify in Section \ref{sec:RQ4}.

\textbf{Humans}. The top three features in Fig. \ref{fig:feature_importance} suggest that human accounts typically use credible sources such as "web" (i.e., Twitter website) and the official Twitter mobile apps. Next, the $account\_age$ and $isGeoEnabled$ features suggest that human accounts have lived relatively long in Twitter and usually have his/her tweets' location enabled, respectively. Also, high timing diversity (entropy) of the tweet, retweet and mention activities are indicative of human accounts, although it is not as high as that of spam bots. Again, Section \ref{sec:RQ4} analyzes this further. Lastly, the $media\_median$ and $media\_mean$ features suggest that human accounts like to attach media files (e.g., photos) in their tweets.

\vspace{-2mm}
\subsection{What Can We Tell about Bots in a Larger Twitter Population?}
\label{sec:RQ4}

To address this question, we performed Experiment $E_2$ by deploying our trained LR classifier to predict for the unlabeled $158,111$ accounts. We then picked the top $K$ accounts with the highest probability scores for each class, and manually assessed the class assignments of these accounts. The assessment results can be found in Appendix \ref{sec:predictions} (Table \ref{tab:top_pred_sg}). We found that the prediction results generally match well with our manual judgments. Based on this, we can make inference on the behavior of bots in a larger Twitter population, i.e., the entire population of Singapore Twitter users. We focus our analyses on the entropy-based dynamic tweet features, which dominate the top features as shown in Fig. \ref{fig:feature_importance}. That is, we analyze the entropy distributions of the tweet, retweet, mention, hashtag and url activities. The complete  distributions can be found in Appendix \ref{sec:predictions} (Fig. \ref{fig:entropy_sg_cdf}), which reveals several interesting insights as elaborated below.

\textbf{Tweet patterns}. We first compared the distributions of the tweet timings, and discovered that consumption and spam bots exhibit higher diversity (entropy) than that of humans. In contrast, broadcast bots were found to have more concentrated timings. These suggest that broadcast bots post tweets at more specific timings than humans and other types of bots. We also found that consumption and spam bots are very similar in terms of daily timings (i.e., weekday and day entropies), but the former is less diverse than the latter in terms of hourly timings. We can thus conclude that consumption and spam bots tweet equally regularly on a daily basis, but the latter tend to post at random hours.

\textbf{Retweet and mention patterns}. Retweet and mention activities can be used to gauge how much a bot (or human) cares about other accounts. Comparing the distributions of the retweet and mention timings in Fig. \ref{fig:entropy_sg_cdf}, we can see again that spam bots have the most random patterns compared to humans and other bot types. But unlike the results for tweet timings, consumption bots have the lowest diversity in terms of daily and hourly timings for the retweet and mention activities. This suggests that consumption bots reshare contents and mention other users at more specific timings, respectively. Such regularity makes sense, especially for consumption bots that provide update services to their users, e.g., Unfollowers and Twittascope (cf. Section \ref{sec:RQ2}).

\textbf{Hashtag patterns}. In Twitter, a hashtag can be viewed as representing a topic of interest. As shown in Fig. \ref{fig:entropy_sg_cdf}, humans and consumptions bots have very similar diversities of hashtag timings. It is also shown that spam bots have the most diverse hashtag timings (as expected), whereas broadcast bots exhibit very focused hashtag timings. The latter suggests that broadcast bots tend to talk about different topics at more regular time intervals. This is intuitive, especially if we consider the nature of the account owners of broadcast bots (e.g, news/blogger sites), which aim to disseminate various information on a regular basis.

\textbf{URL patterns}. For the URL timings, we find that in general humans and broadcast bots use URLs at more specific timings than consumption and spam bots. Interestingly, however, we observe that consumption bots exhibit higher diversity in daily timings than spam bots, but the reverse is true for hourly timings. This suggests that consumption bots use URLs on a more regular daily basis than spam bots, but the latter post URLs at more random hours.

\textbf{Comparisons}. It is also interesting to see how our results in Figs. \ref{fig:feature_importance} and \ref{fig:entropy_sg_cdf} put little emphasis on the importance of the follow network features in the classification task. This is different from previous studies on (malicious) bots \cite{Lee2011,Stringhini2010,Wagner2012,Chu2012,Dickerson2014}, whereby the follow features play a key role. We can attribute this to the evolution of bot activities as well as stricter regulations imposed by Twitter (especially for spam bots). Also, to our best knowledge, no attempt has been made in the previous works to infer on a larger population. Thus, our work offers more comprehensive insights on the behavioral traits of bots.

\vspace{-2mm}
\section{Conclusion}
\label{sec:conclusion}

In this paper, we present a new categorization of bots, and develop a systematic bot profiling framework with a rich set of features and classification methods. We have carried out extensive empirical studies to analyze on broadcast, consumption and spam bots, as well as how they compare with regular human accounts. We discovered that the diversities of timing patterns for posting activities (i.e., tweet, retweet, mention, hashtag and url) constitute the key features to effectively identify the behavioral traits of different bot types. 

This study hopefully will benefit social science studies and help create better user services. In the future, we plan to examine the prevalence of our findings across multiple countries, beyond our current Singapore data. We also wish to study information diffusion and user interaction in Twitter with the aid of bots. 

\vspace{4mm}
\scriptsize
\noindent
\textbf{Acknowledgments}. This research is supported by the National Research Foundation, Prime Minister’s Office, Singapore under its International Research Centres in Singapore Funding Initiative.

\bibliographystyle{abbrv}
\bibliography{main}

\newpage
\begin{subappendices}
\normalsize
\renewcommand{\thesection}{\Alph{section}}

\section{Predictions on Unlabeled Twitter Accounts}
\label{sec:predictions}

To facilitate our study on a larger Twitter population, we first examined how well our best classfier (i.e., LR) can predict for unlabeled data that it never sees in the (labeled) CV data. Table \ref{tab:top_pred_sg} summarizes the top $K$ prediction results, whereby we varied $K$ from $10$ to $50$ to verify the robustness of the predictions. For each class, we computed the number of correctly predicted instances ($TP$) as well as precision at top $K$, i.e., $Precision = \frac{TP}{K}$.

As shown in Table \ref{tab:top_pred_sg}, our LR classifier produces fairly accurate and consistent predictions across different $K$ values. With respect to human accounts, our LR classifier achieved perfect $Precision$ for all $K$ values. Unsurprisingly, we can expect that human accounts constitute the largest proportion of the Twitter population, and thus they should be the easiest to classify. We also obtained good results for the broadcast and consumption bots, with precision scores greater than $75\%$ and $95\%$ respectively. On the other hand, we observe rather modest $Precision$ scores for spam bots (i.e., $40$--$47.5\%$). We can attribute this to the insufficient number of instances for spam bots, which form only $\frac{105}{1,613} = 6.51\%$ of our labeled data (cf. Table \ref{tab:label_dist}). This may (again) be due to our data collection procedure that involved popular users as seeds and/or due to our relatively strict criteria for the characterization of spam bot accounts (cf. Section \ref{sec:RQ1}). Nevertheless, the $Precision$ scores of $40$--$47.5\%$ remain relatively good, if we compare with that of a random guess for our labeled data (i.e., $6.51\%$).

\begin{figure}[!t]
\centering
\includegraphics[width=1.0\textwidth]{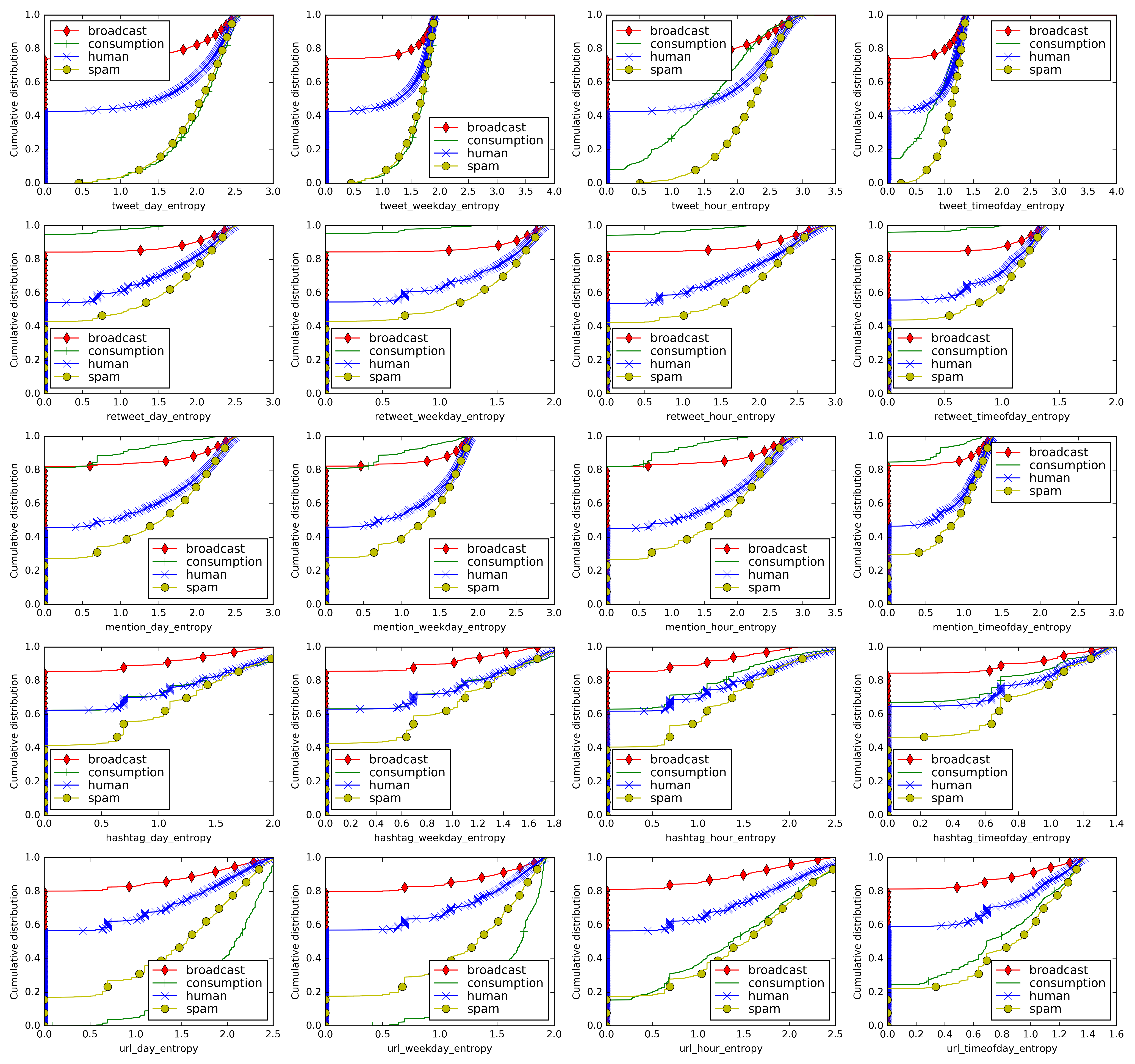}
\caption{Distribution of entropy-based features for 158,111 Twitter accounts}
\label{fig:entropy_sg_cdf}
\end{figure}

All in all, we find our top $K$ predictions on unlabeled data to be satisfactory. Based on this, we can use our predictions to infer the behavioral profiles of bots in a larger Twitter population, which in this case spans the overall Singapore users. In particular, we analyze the entropy-based dynamic tweet features, namely the entropy distributions of the tweet, retweet, mention, hashtag and url activities, which constitute the majority group of the top discriminative features in Fig. \ref{fig:feature_importance}. Fig. \ref{fig:entropy_sg_cdf} presents the cumulative distribution functions of these features. The detailed analysis of the distributions can be found in Section \ref{sec:RQ4}. 

\begin{table*}[!t]
\scriptsize
\centering
\caption{Top $K$ predictions on unlabeled 158,111 Twitter accounts}
\begin{tabular}{|l|c|c|c|c|c|c|c|c|c|c|}
\hline
&\multicolumn{2}{|c|}{$K=10$} & \multicolumn{2}{|c|}{$K=20$} & \multicolumn{2}{|c|}{$K=30$} & \multicolumn{2}{|c|}{$K=40$} & \multicolumn{2}{|c|}{$K=50$}\\
\cline{2-11}
\textbf{Label}	& \textbf{TP} & \textbf{Precision} & \textbf{TP} & \textbf{Precision} & \textbf{TP} & \textbf{Precision} & \textbf{TP} & \textbf{Precision} & \textbf{TP} & \textbf{Precision} \\
\hline
Broadcast bot   & 9  & 0.80    	 & 18 & 0.90      & 27 & 0.90      & 34 & 0.85      & 38 & 0.76\\
Consumption bot & 10 & 1.00      & 20 & 1.00      & 30 & 1.00      & 38 & 0.95      & 48 & 0.96\\
Spam bot        & 4  & 0.40      & 9  & 0.45      & 12 & 0.43      & 19 & 0.475     & 23 & 0.48\\
Human           & 10 & 1.00      & 20 & 1.00      & 30 & 1.00      & 40 & 1.00      & 40 & 1.00\\
\hline
\multicolumn{11}{l}{TP: number of true positives}
\end{tabular}
\label{tab:top_pred_sg}
\end{table*}

\end{subappendices}

\end{document}